\documentclass[a4paper,fleqn,usenatbib,useAMS]{mnras}

\usepackage{graphicx}
\usepackage{amssymb}
\usepackage{subfigure}
\usepackage{captcont}
\usepackage{multirow}
\usepackage{longtable}
\usepackage{comment}
\usepackage{array}
\usepackage{url}
\usepackage{pdflscape}
\usepackage{booktabs,multirow}
\usepackage{times} 
\usepackage{units}
\usepackage{aas_macros} 

\usepackage[T1]{fontenc}
\usepackage{aecompl}
\usepackage{newtxtext,newtxmath}

\usepackage[usenames]{color}

\usepackage{upgreek}

\newcommand{\cd}{d$^{-1}$}

\newcommand{\Dnu}{$\Delta\nu$}
\newcommand{\Msun}{M$_\odot$}
\newcommand{\Rsun}{R$_\odot$}

\title[$K2$ observations of 33\,Lib]{$K2$ observations of the rapidly oscillating Ap star 33\,Lib (HD\,137949): new frequencies and unique non-linear interactions}
\author[D. L. Holdsworth et al.]{Daniel L. Holdsworth,$^{1,2}$\thanks{E-mail: dlholdsworth@uclan.ac.uk}
M. S. Cunha,$^{3}$
H. Shibahashi,$^{4}$
D. W. Kurtz,$^{1}$ and 
\newauthor
D. M. Bowman$^{5}$\\
$^{1}$ Jeremiah Horrocks Institute, University of Central Lancashire, Preston PR1 2HE, UK\\
$^{2}$ Department of Physics, North-West University, Mafikeng Campus, Private Bag X2046, Mmabatho 2745, South Africa\\
$^{3}$ Instituto de Astrof\'isica e Ci\^encias do Espa\c co, Universidade do Porto, CAUP, Rua das Estrelas PT4150-762 Porto, Portugal\\
$^{4}$ Department of Astronomy, The University of Tokyo, Tokyo 113-0033, Japan\\
$^{5}$ Instituut voor Sterrenkunde, KU Leuven, Celestijnenlaan 200D, B-3001 Leuven, Belgium
}
\begin{document}

\date{\today}

\pagerange{\pageref{firstpage}--\pageref{lastpage}} \pubyear{2018} 

\maketitle

\label{firstpage}

\begin{abstract}
We present the analysis of $K2$ short cadence data of the rapidly oscillating Ap (roAp) star, 33\,Librae (HD\,137949). The precision afforded to the $K2$ data allow us to identify at least 11 pulsation modes in this star, compared to the three previously reported. Reoccurring separations between these modes leads us to suggest a large frequency separation, \Dnu, of $78.9\,\muup$Hz, twice that reported in the literature. Other frequency separations we detect may represent the small frequency separation, $\delta\nu$, but this is inconclusive at this stage due to magnetic perturbation of the frequencies. Due to the highly non-linear pulsation in 33\,Lib, we identify harmonics to four times the principal frequency. Furthermore, we note a unique occurrence of non-linear interactions of the 11 identified modes. The frequency separations of the modes around the principal frequency are replicated around the first harmonic, with some interaction with the second harmonic also. Such a phenomenon has not been seen in roAp stars before. With revised stellar parameters, linear non-adiabatic modelling of 33\,Lib shows that the pulsations are {\it{not}} greater than the acoustic cutoff frequency, and that the $\kappa$-mechanism can excite the observed modes. Our observations are consistent with 33\,Lib having a rotation period much larger than 88\,d as presented in the literature. 
\end{abstract}

\begin{keywords}
asteroseismology -- stars: chemically peculiar -- stars: magnetic field -- stars: oscillations -- stars: individual: 33\,Lib -- techniques: photometric.
\end{keywords}

   
\section{Introduction}
\label{sec:intro}

There exists a rare subclass of the chemically peculiar A stars which shows rapid oscillations in short cadence photometric and spectroscopic observations. These stars, known as roAp stars, were discovered by \citet{kurtz82} through targeted photometry of a selection of Ap stars. Since then, only 61 of these objects have been discussed in the literature \citep[for catalogues see][]{smalley15,joshi16}.

The chemical peculiarities in the Ap stars are a result of radiative elevation of, most significantly, singly and doubly ionised rare earth elements in the presence of a strong, stable, global, magnetic field \citep[up to about 30\,kG, e.g.][]{babcock60} which suppresses convection. Typically, but not always, the radiatively elevated elements form chemical spots in the atmosphere of the star around the magnetic poles, and can show abundances of some elements greater that one million times solar \citep{ryabchikova04}. Due to the high stability of the magnetic field, these spots are also stable which can allow for the rotation period of the star to be measured through modulation of its mean light curve. A further property of the magnetic field in Ap stars is that it is misaligned with the rotation axis leading to an observed variable magnetic field strength as the star rotates. This model of understanding the observations of Ap stars is known as the oblique or rigid rotator model \citep{stibbs50}. Finally, the strong magnetic field in the Ap stars is thought to be the reason for their slow rotation when compared to their non-magnetic counter parts \citep{abt95,stepien00}; rotation periods of a few of days to decades or centuries are not uncommon \citep{mathys15}.

The pulsations in the rapidly oscillating Ap stars are apparent in the period range $5-24$\,min with amplitudes up to 34\,mmag in $B$-band observations \citep{holdsworth18a}. Their variability is thought to be driven by the $\kappa$-mechanism acting in the H\,{\sc{i}} ionisation zone causing the excitement of high-overtone pressure modes \citep[p modes;][]{balmforth01,saio05}. However, work by \citet{cunha13} has shown that turbulent pressure is also a viable driving mechanism, especially for those roAp stars pulsating with frequencies above their theoretical acoustic cutoff frequency. The ever increasing number of roAp stars with pulsations higher than the cutoff frequency \citep[see figure 12 of ][for cases]{holdsworth18b} shows a need for an improved understanding of the driving mechanisms in the roAp stars.

One of the goals of asteroseismology is the determination of fundamental stellar parameters through precise stellar models. This is achieved through the measurement of the large and/or small frequency separations (\Dnu\, and $\delta\nu$, respectively). To first order, the asymptotic relation for high-order p modes, which applies to the roAp stars, is given by \citet{tassoul80} as $\nu_{n\ell}\approx\Delta\nu_0(n+\ell/2+\epsilon)$, where $n$ is the radial overtone of the mode, $\ell$ is the angular degree of the mode, $\Delta\nu_0$ is the inverse of the sound travel time across the stellar diameter and $\epsilon$ is a correction term that depends essentially on the properties of the surface layers,. \citet{gabriel85} expressed \Dnu\, as a function of stellar parameters such that $\Delta\nu = 0.205[GM/R^3]^{1/2}$\,Hz where $G$ is the gravitational constant. However, the application of this relation to the roAp stars is complicated by the presence of a magnetic field. Even a `weak' field of only a few kG can perturb a pulsation frequency by between $10-30\,\muup$Hz \citep[$0.864-2.592$\,\cd;][]{dziembowski96,bigot00,cunha00,saio04,saio05,cunha06}. The \Dnu\, value has been measured for about 15 roAp stars, with varying degrees of agreement with spectroscopic and/or interferometric fundamental parameters \citep[e.g. ][]{martinez91,mkrtichian05,bruntt09,sachkov11,perraut13}.

In this paper we present our analysis of {\it{Kepler/K2}} observations of the well known roAp star 33\,Librae. 

 
 \section{33\, Librae (HD\,137949)}
 \label{sec:33Lib}
 
33\,Librae  is a bright ($V=6.69$), well studied roAp star.  We present a compilation of 33\,Lib's properties in Table\,\ref{tab:props}. The parameters shown in this table are mostly found in the work of \citet{shulyak13} where a detailed study of several bright Ap stars was conducted. We note that \citet{kochukhov06} also presented parameters of many chemically peculiar stars, but derive their results through a homogeneous study, rather than from a star-by-star study which is preferable. The radius, and therefore luminosity, provided by \citet{shulyak13} were calculated using the {\it{Hipparcos}} parallax of $11.28\pm0.67$\,mas from \citet{vanleeuwen07}. Given the release of {\it{Gaia}} DR2 data on this star, we provide an updated estimation of the radius and luminosity using the new parallax of $12.48\pm0.06$\,mas \citep{gaia16,gaiaDR2}. We use the updated values throughout the remainder of this paper.

\setlength{\tabcolsep}{2pt}

 \begin{table}
  \caption{Properties of 33\,Lib.}
   \centering
  \label{tab:props}
  \begin{tabular}{lll}
    \hline
    Parameter & Value & Reference \\
   \hline
 Mass (M$_{\odot}$) 		& $1.66\pm0.58$ & \citet{shulyak13}\\
 Radius (R$_\odot$)		& $2.13\pm0.13$ &\citet{shulyak13}\\
 					& $1.92\pm0.16$ & This work$^*$ \\
 Temperature (K) 		& $7400\pm50$   & \citet{shulyak13}\\
 Luminosity (L$_\odot$)	& $12.27\pm1.83$ & \citet{shulyak13}\\
 					& $9.97\pm1.23$	& This work$^*$\\
 Surface gravity ($\log g$)	& $4.0\pm0.1$ 	     & \citet{shulyak13} \\
 Parallax (mas)			& $12.48\pm0.06$ & \citet{gaia16,gaiaDR2}\\
 $\langle B\rangle$ (kG) 	& $4.67\pm0.03$ & \citet{mathys17}\\
 Rotation period (d)		& $7.0187$ & \citet{romanyuk14}\\
 					& $\sim5195$ & \citet{mathys17} \\
 \hline
 \multicolumn{3}{l}{$^*$Scaled from \citet{shulyak13} results using the {\it{Gaia}} DR2 parallax.}\\
 \end{tabular}
 \end{table}
 
 \setlength{\tabcolsep}{6pt}

 Observations of 33\,Lib presented by \citet{kurtz82} revealed the presence of only one pulsation mode, with a period of 8.27\,min ($\nu=174.08$\,\cd). However later observations made in 1987 provided enough evidence for \citet{kurtz91} to suggest a second frequency at about $170.66$\,\cd\, and also gave the first detection of the harmonic to the principal frequency. In discovering this second frequency, \citet{kurtz91} suggested that the large frequency separation, the frequency difference between modes of the same angular degree but consecutive overtone, in this star is about 40\,$\muup$Hz.

Later, \citet{hatzes99} and \citet{mkrtichian03} observed 33\,Lib spectroscopically for the first time with the 2-d coud\'e spectrograph at the McDonald Observatory. They found that, as with other roAp stars, the pulsation radial velocity (RV) amplitude varied with spectral region and line strength. Furthermore, \citet{mkrtichian03} found at least one radial node in the upper atmosphere of 33\,Lib; a conclusion drawn from the anti-phase relation between pulsations measured predominantly in Nd\,{\sc{ii}} and Nd\,{\sc{iii}}. Those findings were confirmed by \citet{kurtz05c} and \citet{ryabchikova07}, with \citeauthor{kurtz05c} detecting a further frequency in spectroscopic observations at 152.84\,\cd. An attempt to verify that new frequency in photometry was presented by \cite{kurtz05b} but to no avail.

Through a spectroscopic study of 10 roAp stars, \citet{kochukhov07} highlighted 33\,Lib as atypical for showing double wave RV variations in its pulsation frequency and a significant amplitude in its harmonic. Our photometric observations may provide a reason for this atypical behaviour.

\citet{sachkov11} performed an analysis of spectroscopic data spanning over 5\,yr. They were able to confirm the second frequency at $170.974\pm0.005$\,\cd and find a different third frequency at $155.75\pm0.01$\,\cd. With the inclusion of this third frequency, the authors claimed to have a ``perfect'' solution to the 5-yr RV curve suggesting mode stability.

Most recently, \citet{ofodum18} presented the results of a 39\,h $B$-band photometric campaign of 33\,Lib from 2013. They suggested there is significant amplitude modulation on a night-by-night basis, however they do not have the resolution to resolve closely spaced modes to discount beating. Analysis of their full data set showed a significant decrease in the amplitude of the principal mode when compared to the results of \citet{kurtz82} and \citet{kurtz91}, which could be indicative of a long period rotational modulation. Finally, they presented a new frequency in 33\,Lib at $173.9233\pm0.0012$\,\cd, a frequency split from the principal by $0.1578\pm0.0013$\,\cd. They proposed this second frequency could be a rotationally split side lobe, suggesting a rotation period of $6.34\pm0.05$\,d.

Most of the detailed work published on 33\,Lib is a result of spectroscopic observations. It is well known that spectroscopic observations of roAp stars are able to detect lower amplitude modes than photometric campaigns as a result of the atmospheric location of the pulsation. However, \citet{holdsworth16b} showed that $K2$ observations have the necessary precision to detect low amplitude pulsation modes usually only seen spectroscopically. This provides us with confidence that the data presented in this work will be the definitive data set of 33\,Lib for many years.

   
\section{$K2$ observations}
\label{sec:K2}

33\,Lib was observed by the {\it{Kepler}} Space Telescope \citep{borucki10} during campaign 15 of its $K2$ mission \citep{howell14} in the short cadence (SC) mode. The campaign started on 2017 August 23 and ended 2017 November 19, covering a total of $88.02$\,d. 

\subsection{Data reduction}

To construct the light curve, we retrieved the target pixel file from the MAST server and created our own custom mask for the target following the method of \citet{bowman18}. Our mask is larger than the standard to account for spacecraft motion and the target moving on the CCD. Although this has the disadvantage of potentially contaminating the light curve with background stars, 33\,Lib is by far the brightest star in the frame at over 2\,000 times brighter than the second brightest star within a 50\,arcsec radius according to the {\sc{simbad}} database. Once created, we used the custom mask to extract the light curve using {\sc{kepextract}} routine from the {\sc{pyke}} tools \citep{still12}. 

Due to the nature of the $K2$ mission, the light curve is dominated by the drift of the telescope which is corrected for every $\sim 6.5$\,h. In an attempt to remove these signatures, we employed the {\sc{kepsff}} task in the {\sc{PyKE}} tools which implements the technique of \citet{vanderburg14} to correct for the motion systematics of the spacecraft. This successfully reduced the effect, but did not completely remove it. To arrive at our final data set, we cleaned the light curve of obvious outliers which were uncorrected by the {\sc{kepsff}} task, and then iteratively prewhitened the light curve to remove most of the low-frequency noise. Our final science light curve consists of 118\,728 points.

Our treatment of the low-frequency signals also removes any astrophysical information in this frequency range too, i.e. signals from rotational variations or low-frequency pulsation modes. However, 33\,Lib is a well studied star with an estimated rotational period of, in most cases, several years \citep{mathys97,mathys17}. We show, in Table\,\ref{tab:props}, another estimate of the rotation period of $7.0187$\,d by \citet{romanyuk14} and find that \citet{wolff75} suggested a period of $23.26$\,d. Neither of these measurements include an error and as such we are dubious of their validity. Most recently, \citet{ofodum18} suggested a rotation period of $6.34\pm0.05$\,d, however this was conjecture based on the possible detection of one rotationally split side lobe to the pulsation. Furthermore, our later pulsation analysis is in favour of the longer estimates of the rotation period for 33\,Lib.

\section{Pulsation analysis}
\label{sec:pul_analysis}

\begin{figure}
\includegraphics[width=\linewidth]{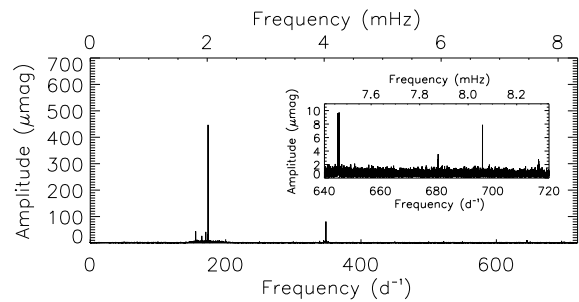}
\caption{Amplitude spectrum of the full frequency range of the $K2$ SC data. The principal pulsation frequency is evident, with further, low-amplitude, modes in the same frequency range. The first harmonic of the principal peak is also visible at this scale. The insert shows a zoom of the region between $640-720$\,\cd\, where further high-frequency peaks are present. These peaks are discussed in the text.}
\label{fig:K2_ft}
\end{figure}

To investigate the full range of possible pulsational variability of 33\,Lib, we calculate a Fourier transform to the Nyquist frequency of the data set, namely $722.735$\,\cd. The result of this is shown in Fig\,\ref{fig:K2_ft}. Clearly evident in the amplitude spectrum is the  principal peak at $174.075$\,\cd, and its first harmonic at twice that frequency. Even at the scale of Fig.\,\ref{fig:K2_ft}, the other pulsations signatures first discovered with spectroscopy are evident around the principal frequency.

Also apparent in Fig.\,\ref{fig:K2_ft} are very high-frequency peaks ($>640$\,\cd). Close inspection of these peaks shows all but one of them to be non-symmetrical and highly jagged, indicating that they are either aliases of frequencies higher than the Nyquist frequency (which we think unlikely) or are not astrophysical in nature. The only `clean' peak occurs at a frequency of $696.2985\pm0.0015$\,\cd\, which is the third harmonic of the principal peak (i.e. $4\nu$). It is not surprising that we have harmonics of the principal mode in this star. The roAp stars are known to show non-sinusoidal variations and, at the precision of {\it{Kepler}} data, the detections of higher harmonics is not uncommon \cite[e.g. ][]{holdsworth16}.

\begin{figure}
\includegraphics[width=\linewidth]{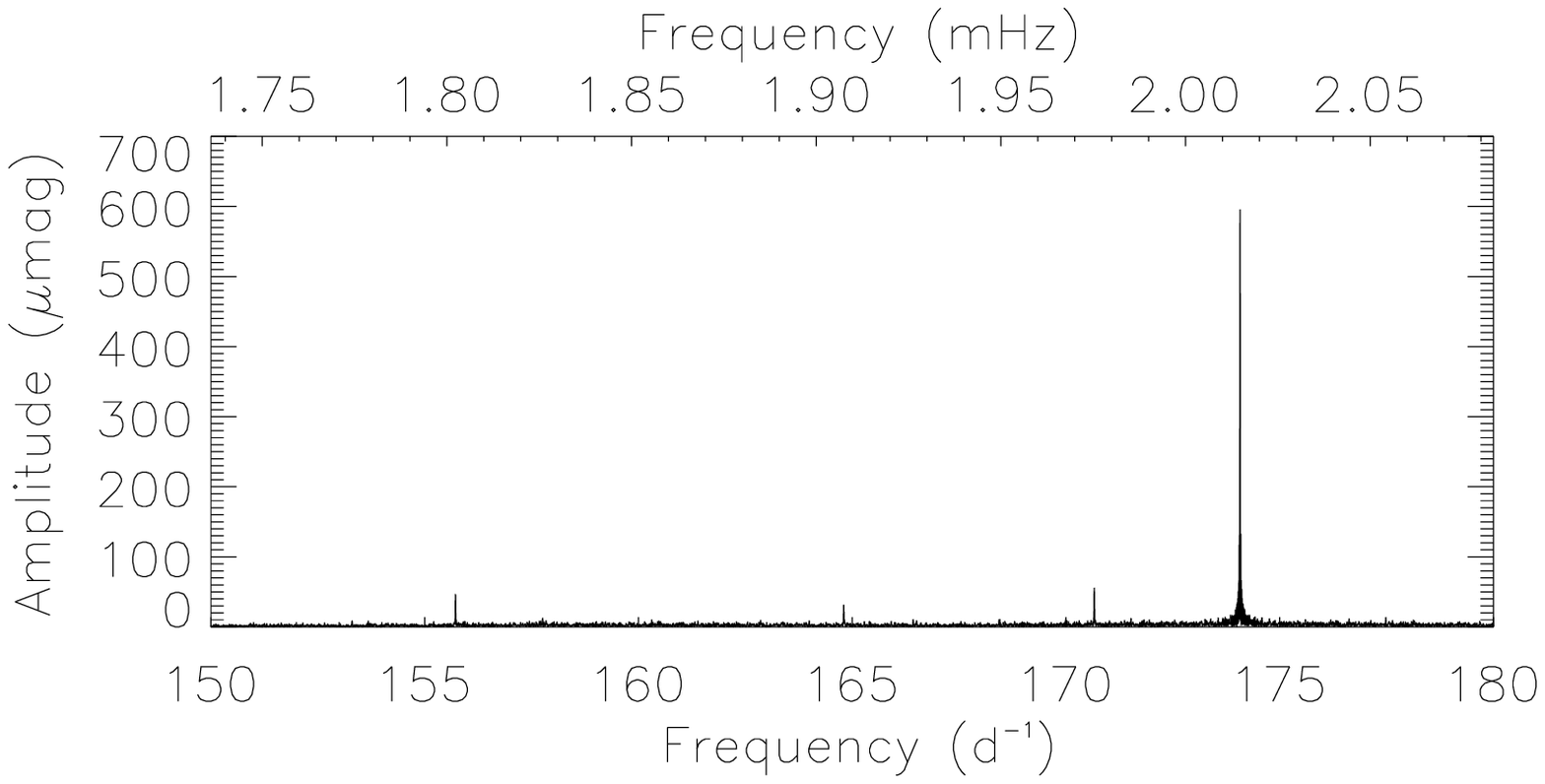}
\includegraphics[width=\linewidth]{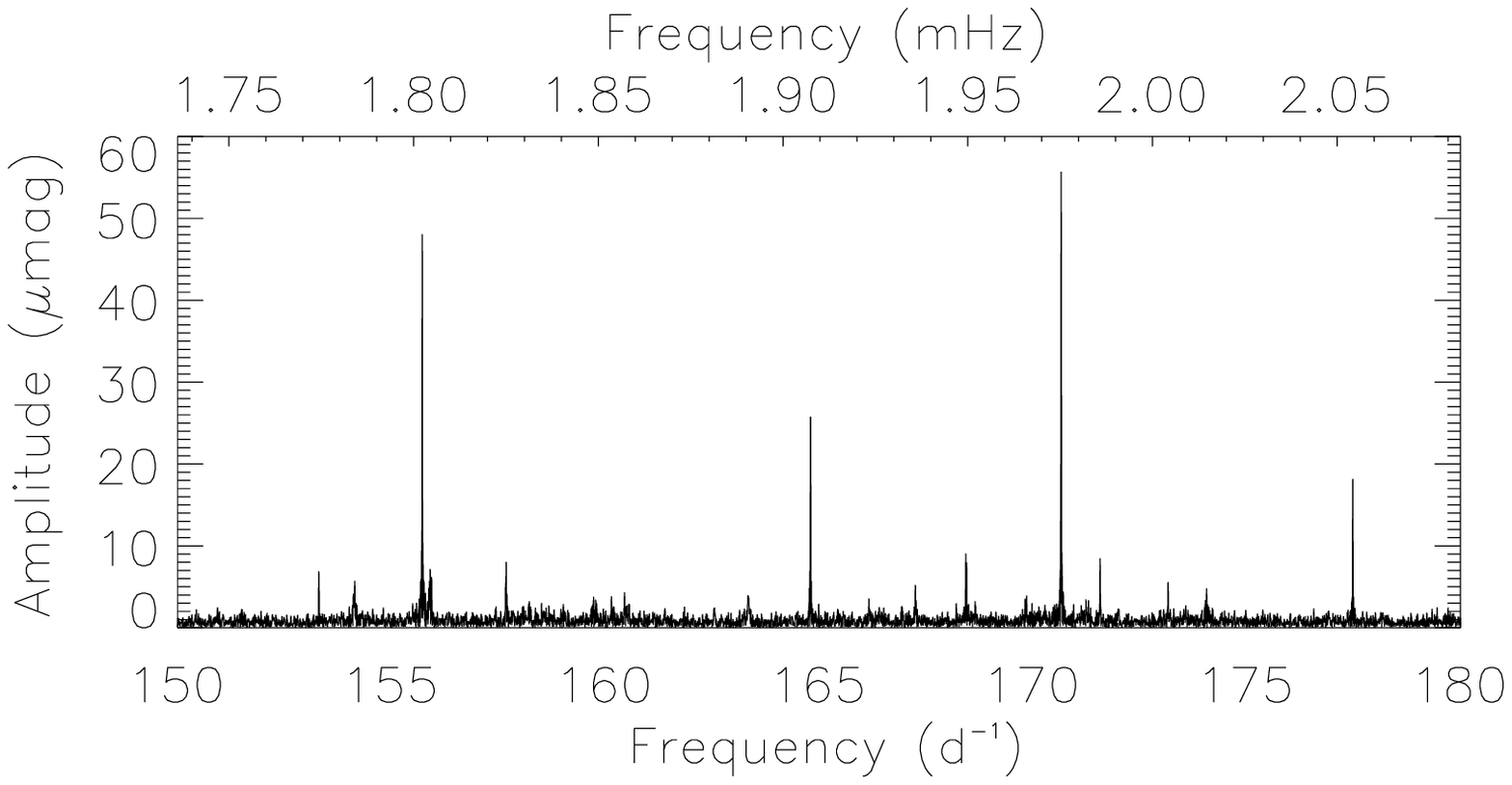}
\caption{Amplitude spectrum of the region around the principal peak. Top: a zoom of the region around the principal pulsation mode. Bottom: enlarged view of the amplitude spectrum after the subtraction of the principal peak. Note the change in scale on the ordinate axis.}
\label{fig:K2_ft_zoom}
\end{figure}

In Fig.\,\ref{fig:K2_ft_zoom} we show a zoomed view of the frequency range around the principal frequency. Evident in the top panel is the principal mode, with a few further low-amplitude modes clearly present. In the bottom panel, we show the amplitude spectrum after the principal peak has been removed which reveals the presence of further modes.

To extract all pulsation frequencies from the light curve we iteratively perform a non-liner least-squares fit to the data and prewhiten each detected frequency. The results of this procedure are shown in Table\,\ref{tab:nlls}. We are able to confidently identify 11 peaks which are significant (i.e. S/N$>4$) and have a `clean' appearance. 

\begin{table}
  \caption{The results of a non-linear least-squares fit to the light curve. The phases are relative to BJD-2458033.8430.}
   \centering
  \label{tab:nlls}
  \begin{tabular}{lcrr}
    \hline
    ID & Frequency & \multicolumn{1}{c}{Amplitude} & \multicolumn{1}{c}{Phase}\\
         & (\cd)      &   \multicolumn{1}{c}{($\muup$mag)}        &  \multicolumn{1}{c}{(rad)}  \\
\hline
 $\nu_1$ & $174.074674\pm0.000006 $ & $ 596.652\pm0.520 $ & $ 1.326\pm0.001$ \\
$\nu_2$ & $170.665270\pm0.000059 $ & $ 55.552\pm0.520 $ & $ 2.082\pm0.009$ \\
$\nu_3$ & $155.722112\pm0.000068 $ & $ 47.976\pm0.520 $ & $ -0.001\pm0.011$ \\
$\nu_4$ & $164.803680\pm0.000128 $ & $ 25.576\pm0.520 $ & $ 1.242\pm0.020$ \\
$\nu_5$ & $177.484091\pm0.000178 $ & $ 18.424\pm0.520 $ & $ -0.772\pm0.028$ \\
$\nu_6$ & $168.435037\pm0.000347 $ & $ 9.463\pm0.520 $ & $ -2.412\pm0.055$ \\
$\nu_7$ & $171.573974\pm0.000397 $ & $ 8.267\pm0.520 $ & $ 1.343\pm0.063$ \\
$\nu_8$ & $157.684610\pm0.000399 $ & $ 8.235\pm0.520 $ & $ -2.179\pm0.063$ \\
$\nu_9$ & $153.304299\pm0.000465 $ & $ 7.058\pm0.520 $ & $ 0.267\pm0.074$ \\
$\nu_{10}$ & $173.165565\pm0.000580 $ & $ 5.655\pm0.520 $ & $ 1.245\pm0.092$ \\
$\nu_{11}$ & $167.254111\pm0.000652 $ & $ 5.036\pm0.520 $ & $ -3.109\pm0.103$ \\

 \hline
 \end{tabular}
 \end{table}
 
\begin{figure}
\includegraphics[width=\linewidth]{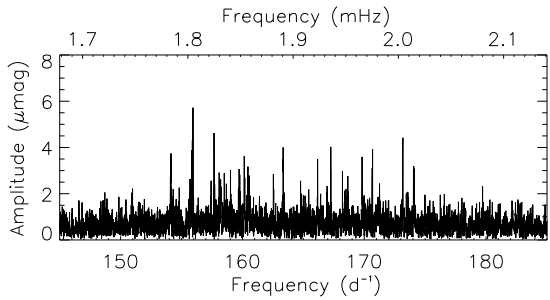}
\caption{The amplitude spectrum after subtracting the frequencies in Table\,\ref{tab:nlls}. The remaining peaks are not `clean' and so have not been extracted.}
\label{fig:aft_sub}
\end{figure}

 Fig.\,\ref{fig:aft_sub} shows the final amplitude spectrum after the subtraction of the frequencies in Table\,\ref{tab:nlls}. There are still some signatures of variability present. However, a combination of their low signal-to-noise and non-clean appearance lead us to note their presence, but not include them in further analysis. 
 
 We do not detect two of the previously published frequencies presented in Sec.\,\ref{sec:intro}, namely the spectroscopically found frequency at $152.84$\,\cd\, from \citet{kurtz05c}, nor the $B$-band photometric frequency at 173.92\,\cd\, found by \citet{ofodum18}. These non-detections could be a result of short mode growth and decay time-scales or energy transfer between modes, as is seen in other roAp stars \citep[e.g.][]{kreidl91,white11}.
 
 The oblique pulsator model \citep{kurtz82, ts94, ts95,bigot02,bigot11}, which describes the pulsations seen in the roAp stars, predicts the presence of sidelobes to the pulsation frequency(ies) that are split by exactly the rotation frequency of the star. This is a result of the varying aspect at which the pulsation poles are viewed, leading to an apparent amplitude modulation of the mode(s). In our data, we do not detect rotational sidelobes to any of the pulsation peaks implying that either (i) the rotation period is longer than the duration of the data set or (ii) our line-of-sight and geometry of the star is such that we view the modes at constant co-latitude (within a few degrees). Although the latter is possible, our $K2$ observations favour the long rotation period hypothesis, discussed below, which supports the claim that the rotation period is longer than the 6.34\,d, 7.018\,d and 23.26\,d found in the literature \citep[][respectively]{ofodum18,romanyuk14,wolff75}.
 
\begin{figure}
\includegraphics[width=\linewidth]{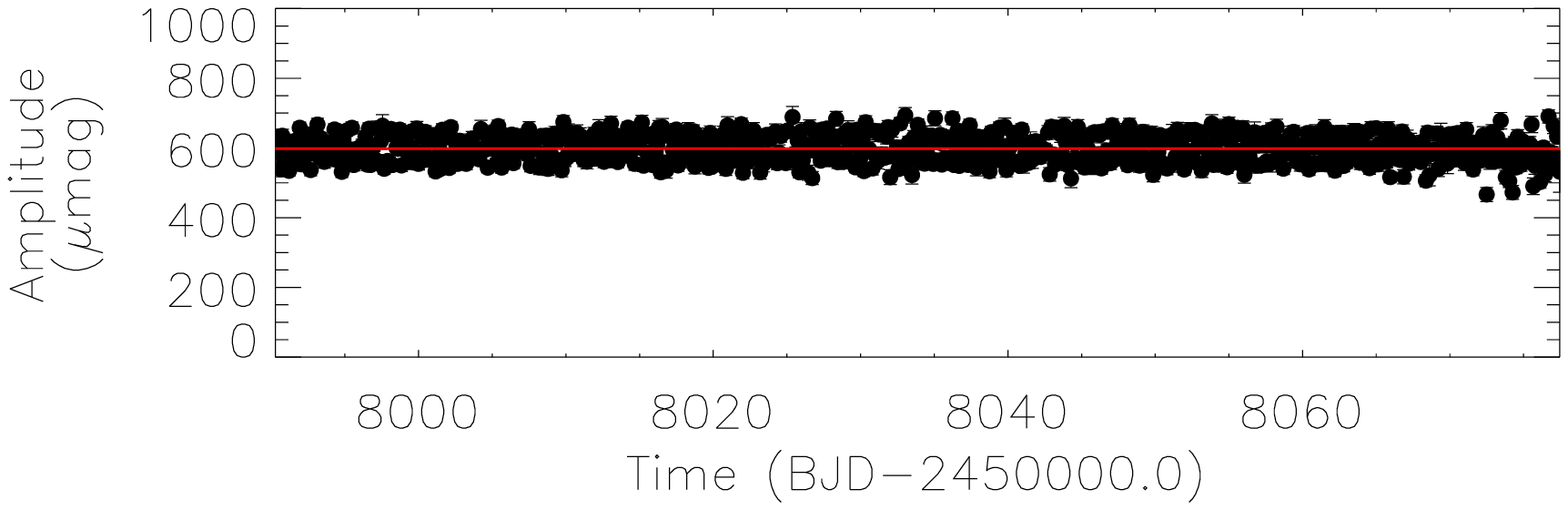}
\includegraphics[width=\linewidth]{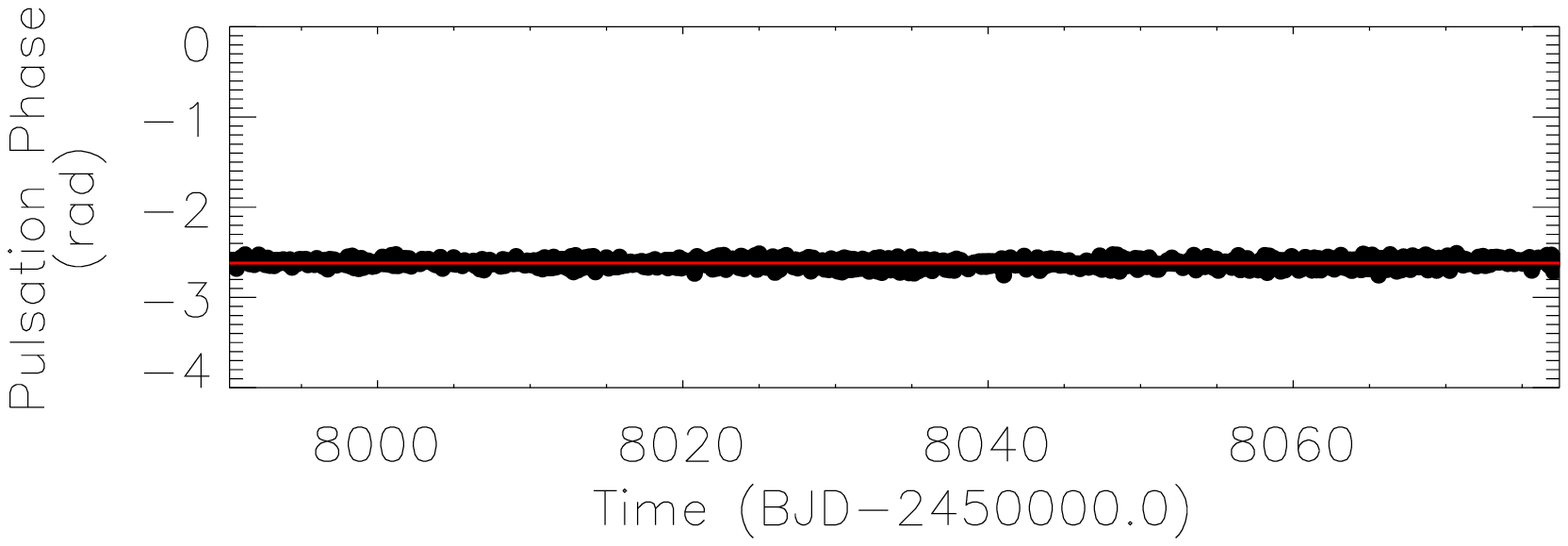}
\caption{The variation of the principal pulsation amplitude (top) and phase (bottom) over the length of observations. At this scale, there is no clear variation in either panel. The red line is to guide the eye only.}
\label{fig:phamp}
\end{figure}

To test the stability of the principal pulsation mode over the observations, we split the light curve into segments of 25 pulsation cycles, or 0.144\,d. We then calculate the amplitude and phase of the pulsation at fixed frequency. This test will provide us with an indication of potential rotational variation seen in the pulsation amplitude, and whether a pulsation node crosses the line-of-sight (as indicated by a phase reversal). The results of the test are shown in Fig.\,\ref{fig:phamp}.
 
The top panel of Fig.\,\ref{fig:phamp} shows no large amplitude variations over the observations, but there is significant scatter about the mean (red) line. The bottom panel also shows no significant change in the pulsation phase of the principal frequency over the length of the observations. These two results demonstrate that a pulsation node does not cross the line-of-sight, and that there is no significant modulation of the mode amplitude over the $K2$ observations in agreement with a long rotation period. Although these results do not rule out the latter case discussed above (i.e. our view of the pulsation mode), we discuss below a possible indication of long term amplitude variations and suggest a test to confirm this.

As 33\,Lib was one of the first roAp stars discovered \citep{kurtz82} there is a long time base of observations to check for amplitude modulation of the principal mode. However, this task is complicated by the different passband of the $K2$ data. Using the conversion between $B$ and $K2$ filters presented in \citet{holdsworth16} for HD\,24355, we suggest that the amplitude measured here would have a $B$ amplitude of $2.59\pm0.28$\,mmag. This is much greater than $1.39\pm0.04$\,mmag presented by \citet{kurtz82} and 1.5\,mmag shown by \citet{kurtz05b}, and a significant increase of the $1.076\pm0.043$\,mmag presented by \citet{ofodum18}. However the same amplitude ratio for HD\,24355 in the different filters may not apply to 33\,Lib. Amplitude ratios vary significantly between different roAp stars \citep{medupe98}, and the broadband {\it{K2}} filter makes comparison with narrowband filter ratios nontrivial. Although an increase in amplitude could be real, new $B$ observations are required to confirm this change. Continued monitoring of the principal mode in 33\,Lib could have the ability to provide the rotation period of this star. 
 
 
\subsection{Frequency separations}

33\,Lib is known to be a multimode pulsating roAp star. \citet{kurtz91} suggested that the frequency separation between $\nu_1$ and $\nu_2$ could be the large frequency separation, i.e. $\Delta\nu\sim40$\,$\muup$Hz. \citet{sachkov11} found a slightly smaller separation between $\nu_1$ and $\nu_2$ of $36\,\muup$Hz. They argued that both this separation and twice its value are inconsistent with what would be expected for the large separation, given the star's global parameters. In fact, considering the parameters from \citet{shulyak13} in Table\,\ref{tab:props} and the scaling of the large separation with the global parameters (see Sec.\,\ref{sec:intro}), we find that the expected \Dnu\, is $\sim53\,\muup$Hz. This is far from the separations presented by \citet{kurtz91} and \citet{sachkov11}, and is also far from twice those values. 

\begin{figure}
\includegraphics[width=\linewidth]{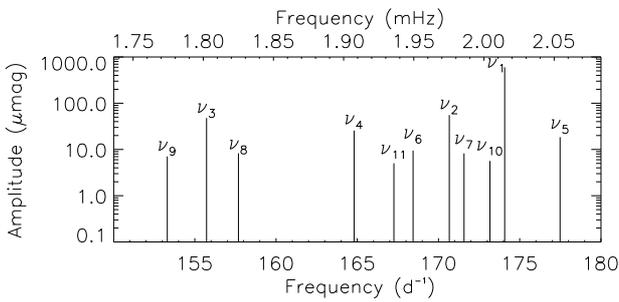}
\caption{Schematic plot of the frequencies shown in Table\,\ref{tab:nlls}. The labels for the frequencies are on the lower frequency side of the peak.}
\label{fig:sepate_fund}
\end{figure}

With the $K2$ photometry, and the presence of many more modes, we revisit the $\Delta\nu$ determination in 33\,Lib. We calculate the differences between all of the frequencies shown in Table\,\ref{tab:nlls}. We note, in Table\,\ref{tab:separations}, frequency differences that reoccur, and the corresponding frequency IDs. For clarity, we also show a labeled schematic plot of these frequencies in Fig.\,\ref{fig:sepate_fund}.
 
 \begin{table}
  \caption{Frequency differences between modes, where the same difference has multiple occurrences. We take the modulus of the difference and quote the highest amplitude peak first. We do not include peaks separated by an integer multiple of the noted separations (i.e. we do not list $\nu_2-\nu_5$).}
   \centering
  \label{tab:separations}
  \begin{tabular}{lcc}
    \hline
    ID & Frequency & Frequency \\
         & (\cd)      &   ($\muup$Hz)    \\
\hline

$\nu_1-\nu_2$ & $3.40940\pm0.00006$ & $39.46066\pm0.00069$ \\
$\nu_1-\nu_5$ & $3.40941\pm0.00018$ & $39.46081\pm0.00207$ \\
$\nu_2-\nu_{11}$& $3.41116\pm0.00065$ & $39.48104\pm0.00758$ \\
\\
$\nu_1-\nu_7$ & $2.50070\pm0.00040$ & $28.94328\pm0.00461$\\
$\nu_2-\nu_{10}$ & $2.50034\pm0.00058$ & $28.93913\pm0.00677$ \\
\\
$\nu_1-\nu_{10}$ & $0.90906\pm0.00058$ & $10.52154\pm0.00674$ \\
$\nu_2-\nu_7$ & $0.90870\pm0.00040$ & $10.51739\pm0.00466$\\

  \hline
 \end{tabular}
 \end{table}
 
 We have easily identified the \Dnu\, value quoted in the literature. Considering that we have four modes separated by $\sim39$\,$\muup$Hz, we suggest that the large separation for 33\,Lib is actually twice the value quoted in the literature, i.e. \Dnu=78.9\,$\muup$Hz. Taking the stellar mass as $1.7$\,\Msun, such a large separation would correspond to a radius of $\sim1.65$\,\Rsun, when applying the scaling from Sec.\,\ref{sec:intro}. This seismic radius is within $2\sigma$ of the radius inferred from the {\it{Gaia}} parallax (see Table\,\ref{tab:props}), but still smaller than that observed.
 
 To test the previously suggested \Dnu\, value of $\sim40\,\muup$Hz, we perform the same calculation and arrive at an expected stellar radius of $2.60\,$\Rsun which is even further way from that derived from the {\it{Gaia}} parallax. We are confident, therefore, that the \Dnu\, value for 33\,Lib is $78.9\,\muup$Hz. 

In this context, it should be recalled that the magnetic field can perturb the frequencies by as much as $10-30\,\muup$Hz \citep{cunha00,saio04,saio05,cunha06}. However, except at particular frequencies, where significant jumps in the magnetic perturbations occur, the perturbation for a given mode degree increases slightly with frequency. Hence the large separations are expected to be less perturbed, increasing by a few $\muup$Hz  at the most, compared to the non-magnetic case. In practice this means that the non-magnetic large separation needed to compute the stellar mean density may be slightly smaller. To obtain the {\it{Gaia}} derived radius (given a mass of $1.70$\,\Msun), the non-perturbed large separation required is $63\,\muup$Hz. This value is further away from the observed value given the magnetic perturbation. If we assume a more reasonable non-perturbed large separation of $74\,\muup$Hz, then we require a minimum stellar mass of $1.80$\,\Msun\, to regain the {\it{Gaia}} derived radius (within the errors), which is consistent with the observations.
 
 Furthermore, there are two other frequencies of note; the separation of $2.50$\,\cd\, (28.94\,$\muup$Hz) and 0.91\,\cd\, (10.52\,$\muup$Hz). From our current observations, we are unsure if either of these frequencies represent the small separation, $\delta\nu$. A value of $\delta\nu=28.94\,\muup$Hz is large for a main-sequence star and is more representative of a zero-age main-sequence star, whereas $\delta\nu=10.52\,\muup$Hz is more applicable to the main-sequence. Previous studies of the small separation in roAp stars have found values around $3-7\,\muup$Hz \citep[e.g.][]{mkrtichian08,saio10}. However, the magnetic perturbations mentioned above will also perturb the small separations. As $\delta\nu$ combines modes of different angular degrees, the perturbation may actually be more significant, and, in practice, both the values observed in 33\,Lib may be plausible. 
  
 One further issue with a large frequency separation of $78.9\,\muup$Hz is that it requires alternating modes of different angular degrees (e.g. $\ell=0, 1, 0, 1$) with almost exact spacing (the difference being on the order of nHz). However, even in the absence of a magnetic field this exact spacing is generally not expected. For example, in the study of $\alpha$\,Cir, \citet{bruntt09} have shown that non-magnetic models may predict an exact spacing only in a specific frequency range. Moreover the magnetic perturbations would be expected to make that spacing even more different, because they would affect modes of different angular degrees differently. Clearly, detailed modelling of 33\,Lib taking into account the magnetic field is needed to identify the observed modes, verify the consistency of model and observed large separation and establish the value of the small separation.

\subsection{Beating}
\begin{figure}
\includegraphics[width=\linewidth]{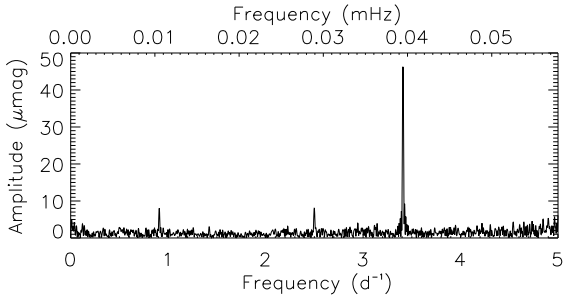}
\caption{Amplitude spectrum of the \emph{variation} of the amplitude of the principal pulsation mode. Clearly present are the re-occurring frequency separations shown in Table\,\ref{tab:separations}. This shows beating is occurring between the modes that are separated by the frequencies in Table\,\ref{tab:separations}}.
\label{fig:phamp_ft}
\end{figure}

\begin{figure}
\includegraphics[width=\linewidth]{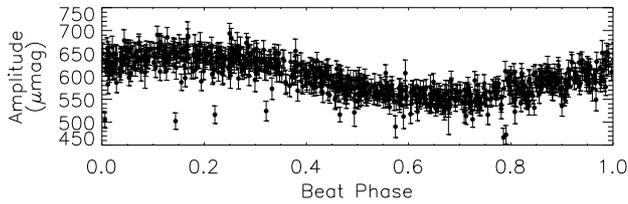}
\caption{The amplitude variation of the dominant peak (i.e. 174.0747\,\cd) folded on the frequency separation of 3.4091\,\cd\, to show the effect of beating between the four frequencies, $\nu_1, \nu_2, \nu_5, \nu_{11}$, separated by this value.}
\label{fig:amp_var_fold}
\end{figure}

In producing the top panel of Fig.\,\ref{fig:phamp}, we notice that there is some deviation from a constant amplitude in this star. Closer inspection of the data used to create Fig.\,\ref{fig:phamp} reveals a periodic variation in the amplitude on a relatively short time scale. The strongest signatures in the amplitude spectrum of this data, shown in Fig.\,\ref{fig:phamp_ft}, are at frequencies of $3.4091\pm0.0041$\cd, $2.4994\pm0.0043$\,\cd\, and $0.9087\pm0.0043$\,\cd. These frequencies correspond to the frequency separations shown in Table\,\ref{tab:separations}. This shows that there is beating of these closely spaced peaks, as one would expect. We fold the amplitude data shown in Fig.\,\ref{fig:phamp} on the dominant frequency (3.4091\,\cd) and show this folded amplitude curve in Fig.\,\ref{fig:amp_var_fold}. The scatter seen in the plot is due to the presence of the two other frequencies.

The beating signal detected here may potentially hamper efforts to measure the rotation period of 33\,Lib through long-term monitoring of the amplitude of its principal pulsation mode. Given a sufficient data length, the effect of beating will be averaged out, which must be considered in any attempt to derive the rotation period in this way.

\subsection{Analysis of the harmonics}

As previously stated, we are able to detect up to the third harmonic of the principal pulsation. For information, we present the extracted harmonics in Table\,\ref{tab:harmonics}.

\begin{table}
  \caption{The harmonics of the principal frequency in 33\,Lib. The zero-point for the phases is the same as that in Table\,\ref{tab:nlls}.}
   \centering
  \label{tab:harmonics}
  \begin{tabular}{lcrr}
    \hline
    ID & Frequency & \multicolumn{1}{c}{Amplitude} & \multicolumn{1}{c}{Phase}\\
         & (\cd)      &   \multicolumn{1}{c}{($\muup$mag)}        &  \multicolumn{1}{c}{(rad)}  \\
\hline
$\nu_1 $ & $174.074673\pm0.000005 $ & $     596.545\pm0.520  $ & $     1.327\pm0.001$\\
$2\nu_1 $ & $348.149428\pm0.000040$ & $     81.099\pm0.520    $ & $     2.040\pm0.006$\\
$3\nu_1 $ & $522.223843\pm0.000460$ & $       7.127\pm0.520    $ & $     2.028\pm0.073$\\
$4\nu_1 $ & $696.298668\pm0.000402$ & $       8.179\pm0.520    $ & $    -2.885\pm0.064$\\
 \hline
 \end{tabular}
 \end{table}

\begin{figure}
\includegraphics[width=\linewidth]{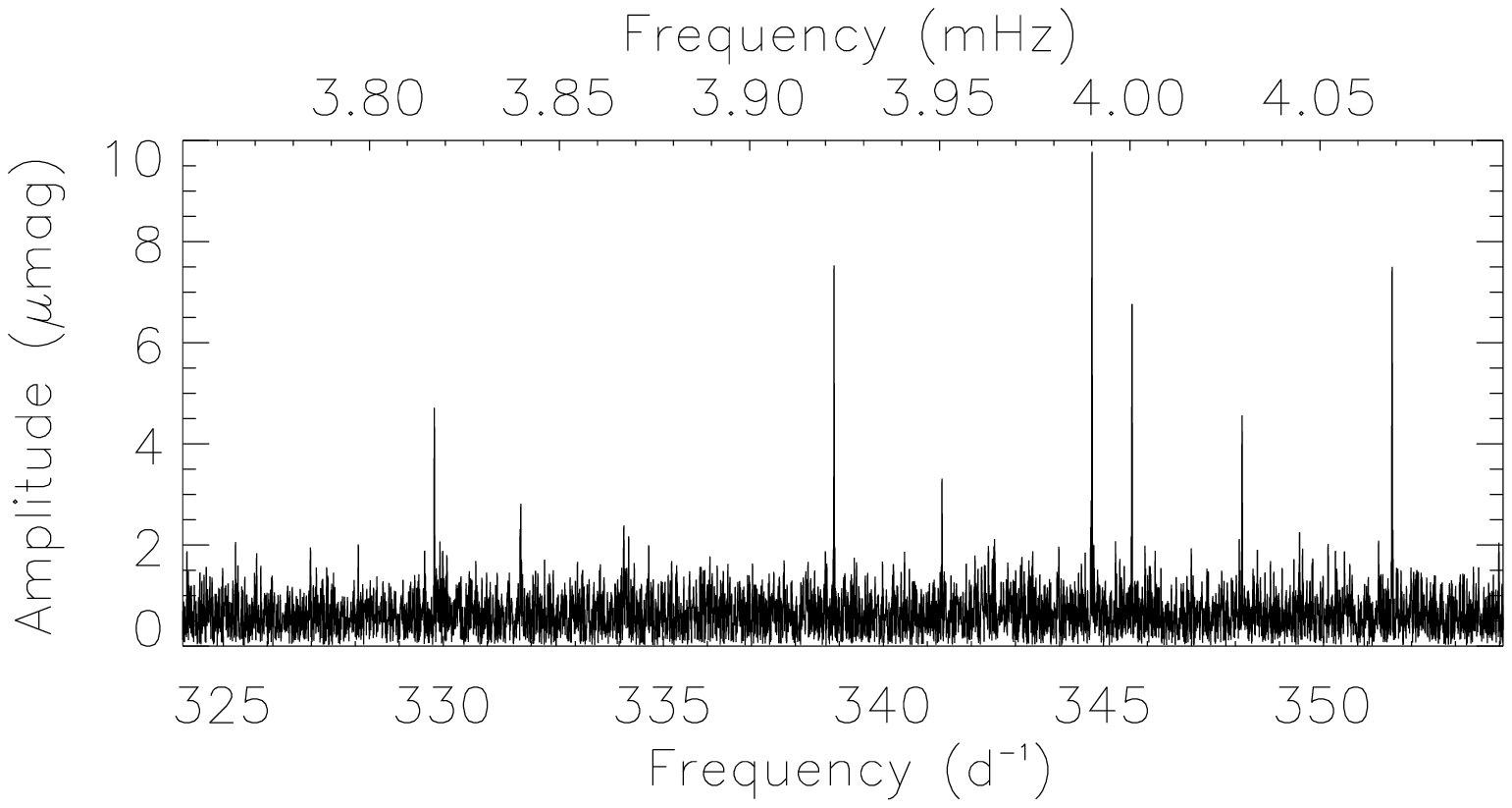}
\includegraphics[width=\linewidth]{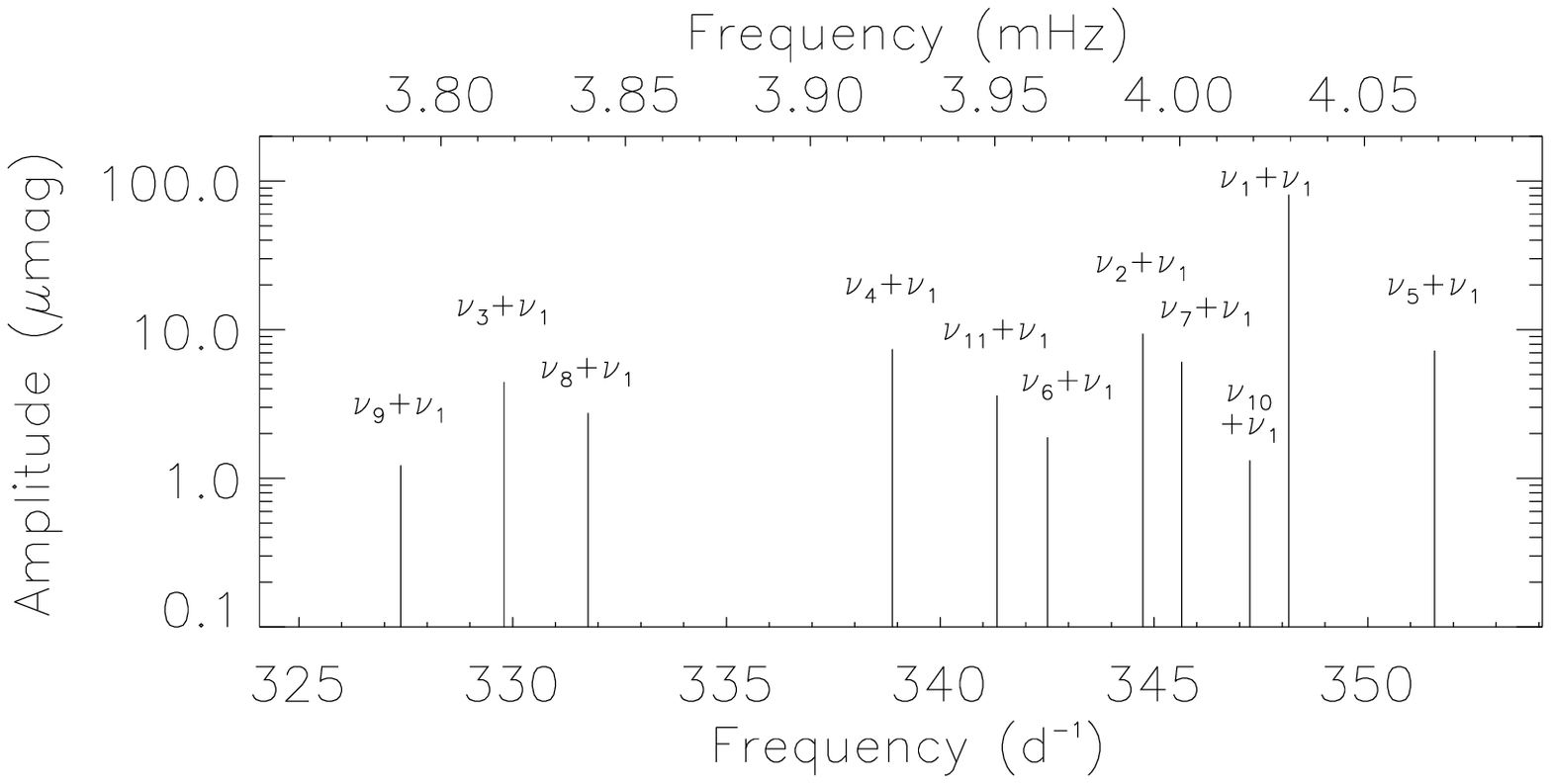}
\caption{Top: amplitude spectrum in the region around the first harmonic, but with the harmonic removed. Note that the separations are the same as for the fundamental. Bottom: schematic plot of the extracted frequencies. }
\label{fig:sepate_harm}
\end{figure}

More interesting than the harmonics themselves is the presence of other peaks in their vicinity. This is most apparent around $2\nu_1$ but also occurs at $3\nu_1$. One would expect that harmonics of the other modes around $\nu_1$ would appear in the amplitude spectrum at two times their frequency. However this is not the case. In Fig.\,\ref{fig:sepate_harm} we show an amplitude spectrum around the first harmonic peak, and a schematic of the extracted frequencies that are presented in Table\,\ref{tab:harm_separ}.

\begin{table*}
  \caption{The frequencies extracted about the first and second harmonics of the principal frequency by non-linear least squares fitting. The zero-point for the phases is the same as that shown in Table\,\ref{tab:nlls}. The final column shows the difference between the fitted value and that expected given the frequency ID. We note that $\nu_9+\nu_1$ and $\nu_{10}+\nu_1$ are below a S/N of 4.0 but include them as they are well recovered under our assumption of non-linear mode coupling.}
   \centering
  \label{tab:harm_separ}
  \begin{tabular}{lcrrr}
    \hline
    ID & Frequency & \multicolumn{1}{c}{Amplitude} & \multicolumn{1}{c}{Phase} & \multicolumn{1}{c}{Calculated difference}\\
         & (\cd)      &   \multicolumn{1}{c}{($\muup$mag)}        &  \multicolumn{1}{c}{(rad)} & \multicolumn{1}{c}{(\cd)}  \\
\hline
 
$\nu_1+\nu_1 $ & $348.149431\pm0.000036 $ & $ 80.966\pm0.456 $ & $ 2.038\pm0.006$ & $ 0.000086\pm0.000036$\\
$\nu_2+\nu_1$ & $344.740731\pm0.000307 $ & $ 9.391\pm0.456 $ & $ 2.415\pm0.049$& $ 0.000790\pm0.000311$\\
$\nu_4+\nu_1$ & $338.878496\pm0.000389 $ & $ 7.394\pm0.456 $ & $ 2.179\pm0.062$ & $ 0.000145\pm0.000405$\\
$\nu_5+\nu_1$ & $351.559364\pm0.000398 $ & $ 7.228\pm0.456 $ & $ -1.398\pm0.063$ & $ 0.000601\pm0.000428$\\
$\nu_7+\nu_1$ & $345.649015\pm0.000473 $ & $ 6.084\pm0.456 $ & $ -0.675\pm0.075$ & $ 0.000371\pm0.000588$\\
$\nu_3+\nu_1$ & $329.796833\pm0.000649 $ & $ 4.431\pm0.456 $ & $ 0.720\pm0.103$ & $ 0.000048\pm0.000652$\\
$\nu_{11}+\nu_1$ & $341.330214\pm0.000797 $ & $ 3.613\pm0.456 $ & $ 2.510\pm0.126$& $ 0.001386\pm0.000976$\\
$\nu_8+\nu_1$ & $331.761371\pm0.001046 $ & $ 2.752\pm0.468 $ & $ 2.085\pm0.170$& $ 0.002121\pm0.001103$\\
$\nu_6+\nu_1$ & $342.508256\pm0.001524 $ & $ 1.890\pm0.456 $ & $ -1.346\pm0.241$ & $ 0.001404\pm0.001554$\\
$\nu_{10}+\nu_1$ & $347.239680\pm0.002192 $ & $ 1.317\pm0.456 $ & $ 1.869\pm0.347$ & $0.000506\pm0.002249$\\
$\nu_9+\nu_1$ & $327.378706\pm0.002362 $ & $ 1.219\pm0.456 $ & $ -1.947\pm0.374$ & $0.000227\pm0.002397$\\
\\
$\nu_5+2\nu_1$& $525.635222\pm0.001664$ & $1.728\pm0.456$ & $0.711\pm0.264$ & $0.001787\pm0.001664$ \\
 \hline
 \end{tabular}
 \end{table*}

We find that the frequencies about the harmonic can all be described by combinations of the principal frequency and the other peaks in that frequency range, suggesting that there is significant non-linear mode coupling occurring in 33\,Lib. To test how closely a combination frequency is to the observed frequency, we calculate the expected frequency and find the difference to the observed frequency. The last column in Table\,\ref{tab:harm_separ} shows the difference between these expected and observed frequencies. 

The agreement, often between $1-2\sigma$, implies that our assumption of non-linear interactions is correct. We perform this check in the `ideal' case, where the frequencies and errors are true representations of the intrinsic frequencies. However the length of the $K2$ observations means that the Rayleigh resolution is no better than about $0.01$\,\cd\, ($0.15\,\muup$Hz) so we argue that all the frequencies in Table\,\ref{tab:harm_separ} are in agreement with the expected results of non-linear mode coupling.

In the case of non-linear interactions, we expect to find the same pattern of frequencies repeated in the low-frequency range too \citep[see e.g.][]{kurtz15,bowman17}, i.e. around $\nu-\nu_1$.  To investigate this, we use the raw \emph{long cadence} $K2$ data (as the noise level is lower) of 33\,Lib which has been subjected to only the {\sc{kepsff}} routine to remove the thruster firings and no iterative pre-whitening of low-frequencies. We find no significant peaks at the expected frequencies however the noise in the this region of $K2$ light curves is significantly higher and may hide the frequencies that we are searching for.  

We pursue this line of enquiry by force fitting, using linear least squares, the expected frequencies to the light curve. As we know the expected frequencies of the peaks, we are able to search in the noise for their presence since the formal amplitude noise is approximately a quarter of the amplitude of the highest peaks. In force fitting a frequency of 3.4094\,\cd, we are able to suggest the presence of a peak at $2.8\sigma$. Although not robustly significant, the presence of the peak here supports our argument of non-linear coupling. Despite the tentative detection of difference $(\nu-\nu_1)$ frequencies, the significant detection of sum $(\nu+\nu_1)$ frequencies detected in 33\,Lib is the first observation of such a phenomenon in the roAp stars.

It is apparent from the amplitudes presented in Table\,\ref{tab:harm_separ} that the precision of the data needs to be in the micro-magnitude range to detect non-linear mode coupling in this star. The most precise ground-based observations of an roAp star achieved a precision of 14\,$\muup$mag \citep{kurtz05} which would not have been sufficient to detect the coupled frequencies seen in 33\,Lib. It is clear that, currently, only space-based observatories are capable of obtaining such precision. Of the multi-periodic roAp stars observed by {\emph{Kepler}}, KIC\,8677585 also shows multiple occurrences of the same separation between some modes around the principal frequency and at the harmonic, which the authors interpret as the large frequency separation \citep{balona13}. There are, however, some frequency patterns around the harmonic which are combinations of the frequencies about the principal, although the authors do not mention (or perhaps notice) this. We propose that the roAp star KIC\,8677585 is also showing non-linear coupling, and the low-frequency peak thought to be a pulsation in this star is actually just a signature of beating between the high-frequency modes.


\section{Modelling}
\label{sec:modelling}

The mechanism responsible for the driving of roAp pulsations has been a matter of debate since these pulsators were first discovered. The instability strip and frequency properties derived based on the excitation by the $\kappa$-mechanism acting on the H\,{\sc{i}} ionization zone in models with envelope convection assumed to be suppressed by the magnetic field \citep{cunha02} can explain the oscillations observed in most of the roAp stars known. However, in some roAp stars, the observed frequencies seem to be too high to be explained by this mechanism, requiring an alternative driving agent, such as the turbulent pressure proposed to explain the pulsations observed in the roAp star $\alpha$\,Cir \citep{cunha13}, whose radius had previously been determined from interferometry \citep{bruntt08}. It is important to note that testing whether the $\kappa$-mechanism may be responsible for the excitation of pulsations in a given roAp star requires an accurate determination of the star's classical parameters, in particular, the stellar radius. An example of this is the roAp star 10\,Aql, whose frequencies were previously thought to be too high to be explained by the $\kappa$-mechanism. Measurement of 10\,Aql's radius by interferometry \citep{perraut13} has, however, demonstrated that the frequency region where this star exhibits pulsations can be reconciled with the region where they are predicted to be excited by the  $\kappa$-mechanism \citep{cunha13}.

33\,Lib is one of the stars that were previously thought to have pulsation frequencies too high to be driven by the  $\kappa$-mechanism. However, the new parallax and consequent downward revision of the star's radius brings that conclusion into question. To test whether this mechanism may be responsible for the observed pulsations we have carried out a linear, non-adiabatic stability analysis based on the star's updated global parameters (see Table\,\ref{tab:props}). The analysis followed closely that performed by \citet{cunha13}. We have considered four different case studies. The first (standard) case is for an equilibrium model with a surface helium abundance of $Y_{\rm surf}=0.01$ and an atmosphere that extends to a minimum optical depth of $\tau_{\rm min}=3.5\times10^{-5}$, and the pulsations are computed with a fully reflective boundary condition. The other three cases are all similar to this one, except that the above options are modified one at the time to: $Y_{\rm surf}=0.1$; $\tau_{\rm min}=3.5\times10^{-4}$; transmissive boundary condition \citep[for further details on the models see][]{cunha13}. In all cases the envelope convection was assumed to be suppressed by the magnetic field at all latitudes, as in \citet{cunha02}, providing the most favourable scenario for excitation.

\begin{figure}
\includegraphics[width=\linewidth]{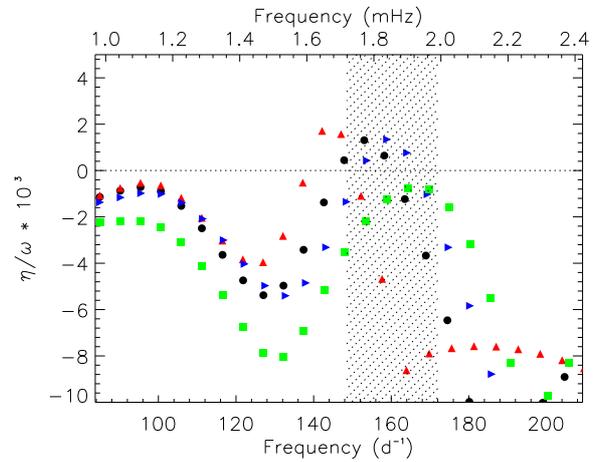}
\caption{Relative growth rates for the four model cases considered. The black circles are for the standard case, right facing blue triangles for the case with $Y_{\rm surf}=0.1$, green squares for the case with $\tau_{\rm min}=3.5\times10^{-4}$, and upwards red triangles for the case with a transmissive boundary condition (see text for details). The diagonally hatched region is where the oscillation frequencies are observed in 33\,Lib. Oscillations are expected when the growth rates, hence the ratio $\eta/\omega$, is positive.}
\label{fig:model}
\end{figure}

Fig.\,\ref{fig:model} presents the growth rates, $\eta$, relative to the real part of the angular oscillation frequencies, $\omega$ as a function of the cyclic frequencies, for the four cases considered. Excitation is expected whenever the growth rates are positive. Also shown is the frequency region where oscillations are observed in 33\,Lib (diagonally hatched region). Overall, the observed frequency range coincides with the frequency range where oscillations are expected to be excited by the $\kappa$-mechanism. We thus conclude that this mechanism can explain the pulsations exhibited by 33\,Lib and emphasise the importance of having accurate stellar parameters before concluding about the need for an alternative driving agent.

\section{Summary and Conclusions}
\label{sec:conclusions}

We have presented the analysis of the most precise data for the well known roAp star 33\,Librae. $K2$ observations over 88\,d have allowed us to identify 11 pulsation modes in this star, far more than previously detected. We also find harmonics of the principal mode up to $4\nu_1$ showing how non-linear the pulsation is in 33\,Lib.

Due to the detection of so many previously unidentified modes, we have been able to find many modes separated by the same frequency. We surmise that one of these frequencies, $3.4094$\,\cd, represents half of the large frequency separation, providing \Dnu$=78.9\,\muup$Hz, which is appropriate for a star on the main-sequence. Of the other two frequency separations, one may represent the small frequency separation, with the other being the difference between the two separations. Confident confirmation of these suggestions will come with detailed, star specific, modelling.

Even after removing the 11 frequencies from the data, the amplitude spectrum (Fig.\,\ref{fig:aft_sub}) still shows some significant variability. We have neglected to fit and remove these signatures as none show a clean sinc-like profile which is perhaps a result of amplitude/frequency variability at these very low amplitudes ($< 7\,\muup$mag). If detailed modelling can predict frequencies that we have not extracted here (Table\,\ref{tab:nlls}), revisiting the peaks in Fig.\,\ref{fig:aft_sub} may be useful.
 
The most intriguing result of this analysis is the repetition of the frequency spacings of the principal peak around its harmonic(s). One would expect that frequencies at the harmonic are separated by twice the value of the separation around the principal frequency. The close to exact spacings suggest that there is significant non-linear mode coupling between the frequencies. We surmise that the harmonics of the frequencies $\nu_2$ through to $\nu_{11}$ have too low amplitudes to be detected, and that the non-linear mode coupling is dominant here. This is the first time that such non-linear interactions have been seen in an roAp star. 

As previously discussed, the oblique pulsator model predicts amplitude modulation of the pulsation mode and side lobes to the mode separated by the rotation frequency. One can argue that we detect both of these model predictions in 33\,Lib; the frequencies split from $\nu_1$ by $3.4094$\,\cd\, could represent a quintuplet expected for a quadrupole mode. The unequal side lobe amplitudes and a missing side lobe at $\nu_1+2\nu_{\rm rot}$ could be a result of either the Coriolis force, unfavourable geometry, mode distortion or a combination of all three. However, we argue here that the separation frequency and the beat period are not representative of the rotation frequency of this star. Such a rotation frequency, given the stellar parameters in Table\,\ref{tab:props}, would mean that 33\,Lib is rotating at $\sim80$\,per cent critical velocity. Such a velocity is unusual for A stars in general, and unheard of in the Ap stars \citep{abt95,royer07}. We are therefore confident in our conclusion that the frequency of $3.4094$\,\cd\, represents ${\Delta\nu}/{2}$, and the amplitude modulation shown in Fig.\,\ref{fig:amp_var_fold} is a result of beating of modes separated by ${\Delta\nu}/{2}$.

Finally, our linear, non-adiabatic modelling of 33\,Lib, with the revised stellar parameters based on the {\it{Gaia}} parallax, has shown that this star is {\it{not}} pulsating above its theoretical acoustic cutoff frequency as previously thought. With a  smaller radius and lower luminosity values, we are able to find models in which oscillations are excited by the $\kappa$-mechanism. This result reinforces the need for accurate stellar parameters when modelling the pulsations observed in the roAp stars.
 
\section*{Acknowledgements}

DLH acknowledges financial support from the Science and Technology Facilities Council (STFC) via grant ST/M000877/1 and the National Research Foundation (NRF) of South Africa.
MSC is supported by Funda\c c\~ao para a Ci\^encia e a Tecnologia (FCT) through national funds (UID/FIS/04434/2013) and by FEDER through COMPETE2020 program (grants POCI-01-0145-FEDER-007672 and POCI-01-0145-FEDER-030389) and through the Investigador FCT contract N$^{\rm o}$\,IF/00894/2012/CP0150/CT0004. 
The research leading to these results has received funding from the European Research Council (ERC) under the European Union's Horizon 2020 research and innovation programme (grant agreement N$^{\rm o}$\,670519: MAMSIE). 
This work has made use of data from the European Space Agency (ESA) mission
{\it Gaia} (\url{https://www.cosmos.esa.int/gaia}), processed by the {\it Gaia}
Data Processing and Analysis Consortium (DPAC,
\url{https://www.cosmos.esa.int/web/gaia/dpac/consortium}). Funding for the DPAC
has been provided by national institutions, in particular the institutions
participating in the {\it Gaia} Multilateral Agreement.
We thank the referee for a careful consideration of the manuscript and complimentary comments and suggestions.

\bibliography{33_Lib-refs}

\label{lastpage}
\end{document}